\documentclass[fleqn,twoside]{article}
\usepackage{gc,graphics,graphicx,cite}
\usepackage{amssymb,amsfonts}

\heads
{K.A. Bronnikov and M.S. Chernakova}
{Charged black holes and unusual wormholes in scalar-tensor gravity}

\def\M{{\mathbb M}}

\def\S{{\mathbb S}}

\def\ME {\mbox{$\M_{\rm E}$}}
\def\MJ {\mbox{$\M_{\rm J}$}}

\def\tg{\tilde{g}}

\def\mn{_{\mu\nu}}
\def\MN{^{\mu\nu}}

\def\Str{\mbox{$\S_{\rm trans}$}}

\def\wh{wormhole}
\def\whs{wormholes}

\def\ssph{static, spherically symmetric}
\def\asflat{asymptotically flat}

\def\BD{Brans-Dicke}
\def\RN{Reissner-Nordstr\"om}

\begin{document}
\twocolumn[
\prepno{gr-qc/0703107}{\GC {13} 51--55 (2007)}

\vspace{1cm}

\Title {Charged black holes and unusual wormholes\yy
	in scalar-tensor gravity}

\Aunames{K.A. Bronnikov\auth{1,\dag,\ddag} and M.S. Chernakova\auth{\ddag}}

\Addresses{
  \addr{\dag} {Centre of Gravitation and Fundamental Metrology, VNIIMS,
        46 Ozyornaya St., Moscow 119361, Russia}
  \addr{\ddag}
    {Institute of Gravitation and Cosmology, Peoples' Friendship University
        of Russia\\
  		  6 Miklukho-Maklaya St., Moscow 117198, Russia}    }

\Abstract
 {We consider \ssph, electrically or/and magnetically charged configurations
  of a minimally coupled scalar field with an arbitrary potential $V(\phi)$
  in general relativity. Using the inverse problem method, we obtain a
  four-parameter family of asymptotically dS, flat and AdS solutions,
  including those with naked singularities and both extreme and non-extreme
  black-hole (BH) solutions. The parameters are identified as the asymptotic
  cosmological constant, an arbitrary length scale, mass and charge. In all
  \asflat\ BH solutions, the potential $V(\phi)$ is partly negative, in
  accord with Bekenstein and Mayo's no-hair theorem. The well-known conformal
  mapping extends the BH solutions to Jordan's pictures of a general class
  of scalar-tensor theories (STT) of gravity under the condition that the
  nonminimal coupling function $f(\phi)$ is everywhere positive. Relaxing
  the latter condition and assuming $f=0$ at some value of $\phi$, we obtain
  \wh\ solutions in a particular subclass of STT. In such solutions, the
  double horizon of an extreme BH in Einstein's picture maps into the second
  spatial infinity in Jordan's. However, at this second infinity, the
  effective gravitational coupling infinitely grows.  }

] 
\email 1 {kb20@yandex.ru}

\section {Introduction}

  Black holes and wormholes are strong field configurations in which
  curvature manifests itself in the global properties of space-time. While
  black holes (BHs) are believed to inevitably result from gravitational
  collapse of sufficiently heavy bodies and are for many years an object (or
  at least a goal) of many astrophysical observations, wormholes have only
  recently appeared in the focus of active discussion. A reason is that
  (traversable, Lorentzian) \whs\ need for their existence (at least in the
  framework of general relativity) some unusual matter able to violate the
  null energy condition $T\mn u^{\mu}u^{\nu}\ge 0$ where $T\mn$ is the
  stress-energy tensor and $u^\mu$ is any null vector.

  Now, with the discovery of the accelerated expansion of the Universe, its
  hypothetic source, the so-called dark energy (DE) seems to be a more or
  less plausible material for \wh\ construction. More precisely, the null
  energy condition is violated if the pressure-to-density ratio $w =p/\eps <
  -1$. By modern observations, $w$ for DE belongs to a range including $-1$,
  but values a little smaller than $-1$ seem to be the most favourable
  \cite{DE-obs}. Various models of DE, their theoretical and observational
  properties are described in the review \cite{copeland}.

  Wormholes are imagined as regular bridges, or tunnels, connecting large or
  infinite regions of space-time, belonging to the same universe or to
  different universes. If they can be sufficiently large and stable, they
  can serve as shortcuts between remote parts of the Universe or as time
  machines. Moreover \cite{KNS06}, they can lead to many observable
  astrophysical effects. It is thus important to know if really existing
  kinds of matter, including DE candidates, can produce and support such
  objects. By now, there are quite a number of various \wh\ solutions, many
  of them obtained with different kinds of phantom matter (e.g., phantom
  scalar fields with negative kinetic energy \cite{br73,h_ell,pha1}), by
  virtue of macroscopic quantum effects \cite{sush,krasn} and in the
  frameworks of generalized, e.g., multidimensional theories of gravity
  \cite{clement,bwh1}; see Refs.\,\cite{wh-rev} for recent reviews. A search
  for realistic \wh\ solutions with more usual and maybe more viable sources
  still remains topical.

  Ref.\,\cite{BrSt07} has studied the possible \wh\ existence in a class of
  scalar-tensor theories (STT) of gravity, in particular, those theories
  which, being non-phantom by nature (i.e., with positive kinetic energy of
  the scalar field in the Einstein picture), are able to produce a
  phantom-like ($w < -1$) behaviour in a certain epoch in cosmology
  \cite{GPRS06}. It has been shown \cite{BrSt07} that, even in the presence
  of electric or magnetic fields, if the non-minimal coupling function
  $f(\Phi)$ is everywhere positive, \wh\ solutions are absent, and this
  conclusion holds in both Einstein and Jordan pictures. It also turned out
  \cite{BrSt07} that if $f$ remains non-negative but is allowed to reach
  zero at some value of the scalar field $\Phi$, \whs\ in Jordan's picture
  are not excluded though require a very specific kind of STT and severe
  fine tuning.

  In this paper, we give explicit analytic examples of such \ssph\ \wh\
  solutions.  To do that, it proves necessary to build an extreme BH
  solution with a scalar field in Einstein's picture. The latter is
  impossible if the scalar field is the only source of gravity, as follows
  from the global structure theorem \cite{vac1} for scalar fields with
  arbitrary potentials in general relativity, but can exist if a radial
  electric or magnetic field is present. To find them, in \sect 2, we
  consider \ssph, minimally coupled scalar (with an arbitrary potential
  $V(\phi)$) and electromagnetic fields in general relativity. Using the
  inverse problem method, we obtain a family of solutions, including those
  with naked singularities and both extreme and non-extreme charged BH
  solutions. They seem to be the first explicit examples of charged \asflat\
  BH solutions with scalar and electromagnetic fields interacting only via
  gravity\footnote
  	{A direct interaction in a dilatonic form, $F(\phi)\, F\MN F\mn$,
	 $F = \e^{\lambda\phi},\ \lambda = \const$, leads to the well-known
	 dilatonic BHs, found for the first time in Ref.\,\cite{br-sh77}.
	 Charged \asflat\ BHs with another form of direct interaction,
	 $F(\phi) = (1 + l^2\phi^2)^{-1}$, were recently discussed by Gubser
	 \cite{gubs05}. Examples of charged hairy asymptotically AdS BHs
	 with $V(\phi) < 0$ were obtained in Ref.\,\cite{mart06}.}.
  In the BH solutions found here, the field potential $V(\phi)$ is partly
  negative, in accord with the no-hair theorem \cite{BekMayo96} stating that
  charged \asflat\ BHs cannot possess scalar fields with non-negative
  potentials outside their event horizons.

  In \sect 3, using the appropriate conformal mappings, we extend the BH
  solutions of \sect 2 to a general class of STT under the condition that
  $f(\phi) >0$ everywhere. Then we obtain and briefly discuss \wh\ solutions
  in some particular, specially designed STT in which $f=0$ at some value of
  $\phi$.  In these solutions, the double horizon of an extreme BH in
  Einstein's picture maps into the second spatial infinity (another mouth of
  the \wh) in Jordan's.

\section{Charged BHs in Einstein's picture}

  Consider general relativity with minimally coupled scalar ($\phi$) and
  electromagnetic ($F\mn$) fields as sources. The Lagrangian is\footnote
	{Our conventions are: the metric signature $(+{}-{}-{}-)$; the
	curvature tensor
	$R^{\sigma}{}_{\mu\rho\nu} = \d_\nu\Gamma^{\sigma}_{\mu\rho}-\ldots,\
	R\mn = R^{\sigma}{}_{\mu\sigma\nu}$, so that the Ricci scalar
	$R > 0$ for de Sitter space-time or a matter-dominated cosmology;
	the system of units $\hbar = c = 1$.}
\bearr \nhq
       L_E = \frac{1}{2}\bigl[R + g\MN \phi_{\mu}\phi_{,\nu}
		    	    - 2V(\phi) - F\MN F\mn \bigr],  \label{LE}
\ear
  where $R$ is the Ricci scalar and $V(\phi)$ is the scalar field potential.

  Let us assume static spherical symmetry, considering the
  theory (\ref{LE}) in a space-time with the metric
\beq                                                            \label{ds}
    ds_E^2 = A(\rho) dt^2 - \frac{d\rho^2}{A(\rho)} - r^2(\rho) d\Omega^2,
\eeq
  where $d\Omega^2 = d\theta^2 + \sin^2\theta\, d\varphi^2$, and putting
  $\phi = \phi(\rho)$. The Maxwell fields compatible with spherical symmetry
  are radial electric ($F_{01}F^{10} = q_e^2/r^4$) and magnetic
  ($F_{23}F^{23} = q_m^2/r^4$) fields, where the constants $q_e$ and
  $q_m$ are the electric and magnetic charges, respectively.

  The scalar field equation and three independent combinations of the
  Einstein equations read
\bear
	 (Ar^2 \phi')' \eql   r^2 dV/d\phi,                 \label{phi}
\\
         (A'r^2)' \eql - 2r^2 V + 2q^2/r^2;                 \label{00E}
\\
              2 r''/r \eql -{\phi'}^2 ;                     \label{01E}
\\
         A (r^2)'' - r^2 A'' \eql 2 -4q^2/r^2,              \label{02E}
\ear
  where the prime denotes $d/d\rho$ and $q^2 = q_e^2  + q_m^2$.
  \eq(\ref{phi}) follows from (\ref{00E})--(\ref{02E}), which, given a
  potential $V(\phi)$, form a determined set of equations for the unknowns
  $r(\rho),\ A(\rho),\ \phi(\rho)$. The scalar field is normal,
  with positive kinetic energy, whence, due to (\ref{01E}), we have
  $r'' \leq 0$, which forbids \wh\ throats (minima of $r(\rho)$), to say
  nothing of \whs\ as global configurations.

  To obtain examples of BH solutions, we will use the inverse problem
  method.  Namely, we choose the function $r(\rho)$ and then, consecutively,
  find $A(\rho)$ from \eq (\ref{02E}), $\phi(\rho)$ from (\ref{01E}) and
  $V(\rho)$ from (\ref{00E}). If $r''< 0$ everywhere, the function
  $\phi(\rho)$ is strictly monotonic, and therefore the potential $V(\phi)$
  obtained from $\phi(\rho)$ and $V(\rho)$ is well-defined.

  So, we make the simple choice
\beq
	r(\rho) = (\rho^2-b^2)^{1/2},                       \label{r}
\eeq
  where $b$ is an arbitrary constant. Then the field equations give:
\bear                                                        \label{sol-A}
	A(\rho) \eql B_0 r^2 + 1 + 3m\biggl[-\frac{\rho}{b^2}
		+\frac{r^2}{2b^3} \ln\frac{\rho+b}{\rho-b}\biggr]
\nnn\qquad
		-\frac{q^2}{b^4} \biggl[b^2-b\rho\ln\frac{\rho+b}
		{\rho-b}+\frac{r^2}{4} \ln^2\frac{\rho+b}{\rho-b} \biggr],
\yy                                                        \label{sol-phi}
	\phi(\rho) \eql \phi_0 + \frac{\sqrt{2}}{2}\ln\frac{\rho+b}{\rho-b},
\yy
	V(\rho) \eql - \frac{B_0 (3\rho^2 - b^2)}{r^2} +
	      \frac{9 m \rho r^2 + q^2 (3\rho^2 -2b^2)}{b^2 r^4}
\nnn\qquad                                                  \label{sol-V}
	- \frac{3m (3\rho^2-b^2) + 6 q^2\rho}{2b^3 r^2}
		\ln \frac{\rho + b}{\rho - b}
\nnn\qquad
	+ \frac{q^2 (3\rho^2-b^2)}{4b^4 r^2}\ln^2 \frac{\rho+b}{\rho-b}.
\ear
  where $B_0$, $m$ and $\phi_0$ are integration constants. As
  $\rho\to\infty$, the metric becomes flat or (anti-)de Sitter according to
  the value of $B_0 = -V (\infty)/3$, where $V(\infty)$ plays the role of a
  cosmological constant at large $\rho$.

  In what follows, to have a flat asymptotic as $\rho \to \infty$, we put
  $B_0 =0$ and also, without loss of generality, take $\phi_0 =0$.
  Then, at large $\rho$, $r \approx \rho$ and
\beq
	A\approx 1 - \frac{2 m}{\rho} + \frac{q^2}{\rho^2} + \ldots,
\eeq
  so that the metric is approximately \RN, and $m$ is the Schwarzschild mass
  (in units of length).

  The other extreme is $\rho\to b$ ($r\to 0$, the singularity), where
\bearr
      r \approx \sqrt{2b(\rho - b)},
   \cm
      A \approx \frac{q^2}{b^2}\ln\frac{2b}{\rho-b} \to \infty,
\nnn
      V \approx q^2/r^4 \to \infty.
\ear
  In the scalar-vacuum case $q=0$, we recover the solution obtained in
  Ref.\,\cite{vac2}, where examples of BHs with a scalar field and a
  Schwarzschild-like global structure were obtained. Note that in this
  solution, though $A$ tends to a finite limit as $\rho\to b$, it is a
  singularity where the potential $V$ is infinite:
\bear
     A \to 1 - 3m/b, \cm
     V \approx \frac{3m}{br^2} \ln\frac{\rho-b}{2b} \to -\infty.
\ear

  Possible horizons are described by zeros of the function $A(\rho)$, and
  for $q\ne 0$, just as in the \RN\ solution, the number of horizons may be
  0, 1 (a double horizon) and 2, which correspond to configurations with a
  naked singularity, extreme BHs and non-extreme BHs, respectively.

  The values of $m$ and $q$ corresponding to a double horizon $\rho = \rho_h$,
  near which $A(\rho) \sim (\rho-\rho_h)^2$, can be found numerically.
  For this calculation let us introduce the dimensionless quantities
\bearr		\label{dimls}
   \ x = \rho/b, \cm  \bar{r}(x) = r/b, 
\nnn
    \bar{m} =m/b, \ \ \ \cm \bar{q} = q/b, \cm\, \bar{V} = Vb^2.
\ear
  In what follows, we will use these quantities omitting the bars. Thus
  all dimensionful quantities are now measured in units equal to appropriate
  powers of the arbitrary length $b$ (recall that due to $c = \hbar =1$ all
  dimensions are represented as powers of length).

  Then, for some values of $m$, the charges $q = q_2(m)$ providing a double
  horizon and the corresponding coordinate values $x = x_h = \rho_h/b$
  are as follows:
\beq                                                        \label{dub}
	\begin{array} {lll}
  	m = 0.35,\ & q_2 \approx 0.08096, \ & x_h \approx 1.00116; \\[3pt]
   	m = 1,   \ & q_2 \approx 0.9222,  \ & x_h \approx 1.324;   \\[3pt]
   	m = 3,   \ & q_2 \approx 2.97409, \ & x_h \approx 3.1171;  \\[3pt]
   	m = 10,  \ & q_2 \approx 9.9922,  \ & x_h \approx 10.036;  \\[3pt]
   	m = 40,  \ & q_2 \approx 39.9981, \ & x_h \approx 40.01.
	\end{array}
\eeq
  Evidently, as $m$ and $q$ grow, these parameters more and more approach
  those of the \RN\ solution. And, in full similarity with the latter, given
  the mass $m$, for $q > q_2(m)$ the system has no horizon while for $q <
  q_2(m)$ it possesses two simple horizons.

  Figs.\,1 and 2 show the behaviour of the functions $A(\rho)$ and $V(\rho)$
  for some values of the solution parameters. Fig.\,2 indicates that, as $q$
  grows, $V(\phi)$ also grows, at least in the range of $x$ shown. This may
  seem to question the validity of the no-hair theorem \cite{BekMayo96} that
  asserts the non-existence of charged BH solutions with $V(\phi) \geq 0$
  for a class of theories including ours as a special case. An inspection
  shows that our BH solutions respect the theorem: at larger $x$ there is
  always a range of $x$ with negative (though tiny) values of $V$, see an
  example in Fig.\,3.

\EFigure{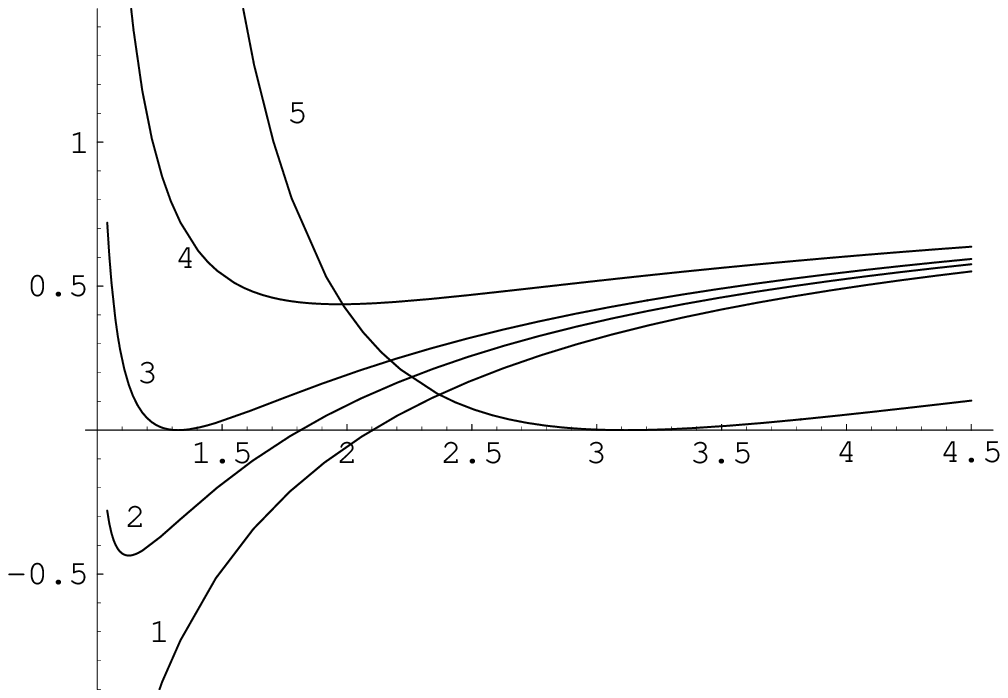}   
    {The metric function $A(x)$ in the solutions (\ref{r})--(\ref{sol-V}).
    The curves correspond to the following parameter values:
     1 --- $m=1$, $q=0$ (scalar-vacuum solution);
     2 --- $m=1$, $q=0.7$ (with two horizons);
     3 --- $m=1$, $q=0.9222$ (with a double horizon);
     4 --- $m=1$, $q=1.3$ (without a horizon);
     5 --- $m=3$, $q=2.97409$ (with a double horizon).}

\EFigure{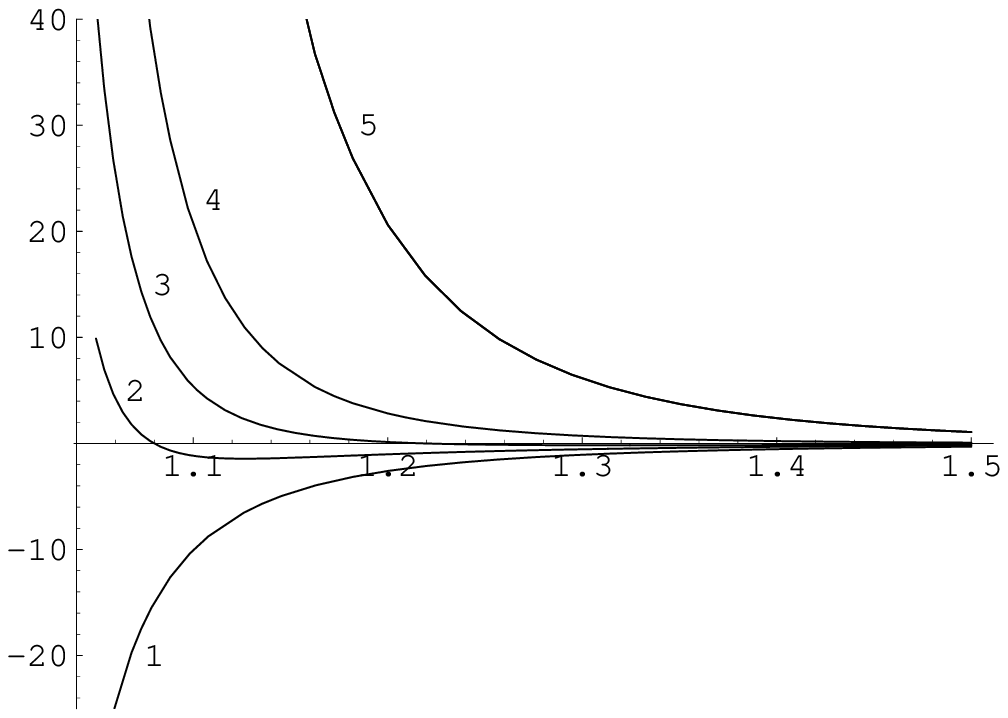}  
    {The potential $V(x)$ in the solutions (\ref{r})--(\ref{sol-V}). The
    curve labels refer to the same parameter values as in Fig.\,1.  }

\EFigure{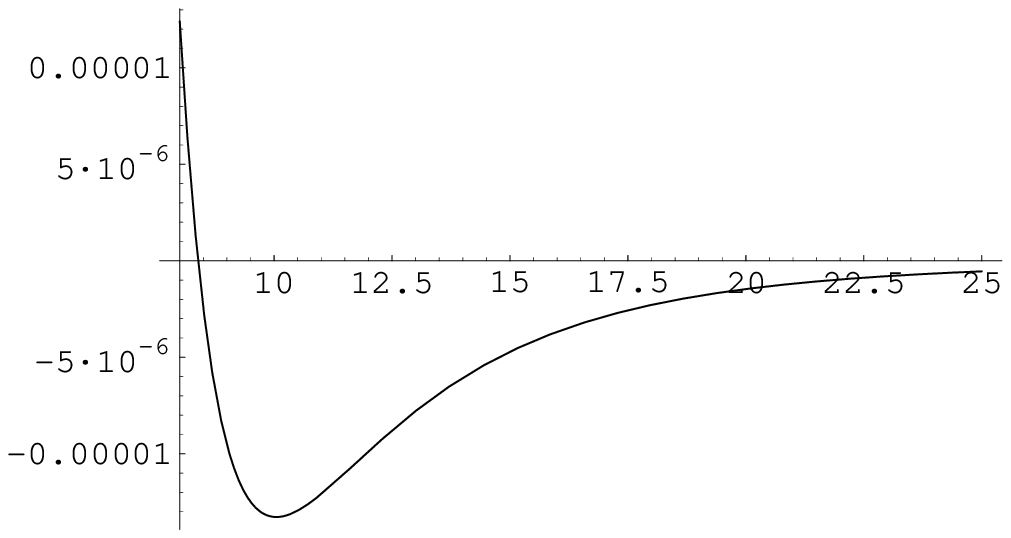}  
    {The potential $V(x)$ for $m=10$, $q=9.99222$ (an extreme BH solution)
    at larger $x$.}

\section {Charged black holes and wormholes in Jordan's picture}

    The Lagrangian of a general STT of gravity in a Jordan-frame manifold
    \MJ\ with the metric $\tg\mn$ and the Maxwell field $F\mn$
    may be written as
\beq   \nq\,
	2L= f(\Phi) {\tilde R} + h(\Phi)\tg^{\mu\nu}\Phi_{,\mu}\Phi_{,\nu}
				- 2U(\Phi) - F\MN F\mn,       \label{LJ}
\eeq
    where ${\tilde R}$ is the Ricci scalar corresponding to $\tg\mn$ while
    $f(\Phi)$ (the nonminimal coupling function), $h(\Phi)$ and $U(\Phi)$
    are arbitrary functions, changing from one particular STT to another.
    Suppose (as is usually done)
\beq                                                          \label{no_ph}
      f \geq 0,
   \cm
      l(\Phi) := fh + \frac{3}{2}\biggl(\frac{df}{d\Phi}\biggr)^2 \geq 0,
\eeq
    then the well-known conformal mapping
\beq
	\tg\mn = f(\Phi) g\mn,                               \label{map}
\eeq
    from \MJ\ to another manifold \ME\ with the metric $g\mn$ converts
    the theory to another formulation called the Einstein frame (or
    picture). The Lagrangian (\ref{LJ}) is transformed (up to a full
    divergence) to (\ref{LE}), where the scalars $\Phi$ and $\phi$ and the
    potentials $U$ and $V$ are connected by the relations
\bearr
    	\frac{d\phi}{d\Phi} = \frac{\sqrt{l(\Phi)}}{f(\Phi)},
    \cm
	V(\phi) = f^{-2} U(\Phi).                           \label{trans}
\ear

    The Lagrangian (\ref{LJ}) admits reparametrization of the field $\Phi$,
    and without loss of generality we may use the so-called \BD\
    parametrization in which
\beq    \nhq                                                  \label{BDpar}
	f(\Phi) = \Phi, \quad\
	h(\Phi) = \frac{\omega(\Phi)}{\Phi}.\quad\
	l(\Phi) = \omega (\Phi) + \frac{3}{2}.
\eeq
    (The \BD\ theory is then the special case $\omega = \const$.)

    Now, any solution to the field equations due to (\ref{LE}), including
    the BH solutions of \sect 2, may be treated as solutions of an
    {\it arbitrary\/} STT (\ref{LJ}) in its Einstein picture. If the
    conformal factor $f$ is everywhere positive, then spatial infinity in
    one picture maps to spatial infinity in another, a horizon maps to a
    horizon of the same order, and a centre to a centre. Therefore our BH
    solutions have counterparts in Jordan pictures of all such STT.
    A good example is the \BD\ theory with $\omega > -3/2$, for which the
    conformal factor is
\beq
	f = \Phi = \exp \left(\phi/\sqrt{\omega + 3/2}\right).
\eeq

    Of interest for us is also the case of a possible conversion of an
    extreme BH in Einstein's picture to a traversable \wh\ in Jordan's. As
    shown in Ref.\,\cite{BrSt07}, this is only possible for such STT that,
    in the parametrization (\ref{BDpar}), $l (\Phi) \sim \Phi$ where $\Phi
    \to 0$ corresponds to a double horizon in \ME\ and to the second spatial
    infinity in \MJ. Evidently, in all such cases one must have $f(\Phi)
    \sim (x-x_h)^2$ as $x\to x_h$ to obtain a finite value of $\tg_{tt}$.

    Let us give an example using the extreme BH from \sect 2 and assuming
    the conformal factor $f$ depending on the coordinate $x$ as
\beq
	f(\Phi) = \Phi = (1 - x_h/x)^2.                        \label{f-wh}
\eeq
    Flat spatial infinity at $x = \infty$ is then provided in both
    manifolds \ME\ and \MJ. The metric in \MJ\ is now
\bear                                                         \label{dsJ}
     ds_J^2 = \frac{x^2}{(x-x_h)^2} \biggl[
	       A(x) dt^2 - \frac{b^2 dx^2}{A(x)} - b^2 r^2 d\Omega^2 \biggr],
\ear
    where $r^2 = x^2 -1$; $A(x)$ and $V(x)$ are given by (\ref{sol-A}) and
    (\ref{sol-V}) with the following substitutions according to our
    asymptotic flatness assumption and transition to dimensionless
    quantities:
\[
     B_0=0, \qquad  b \mapsto 1,\qquad  \rho\mapsto x.
\]
    The STT itself is characterized, in the \BD\ parametrization, by the
    functions
\bearr \nhq                                                    \label{om-wh}
	\omega(\Phi) + \frac{3}{2} = \frac{x^2 (x-x_h)^2}{x_h^2 (x^2-1)^4},
    \quad\
 	U(\Phi) = \frac{(x-x_h)^4}{x^4} V(x),
\nnn
\ear
    where $x = \coth (\phi/\sqrt{2}) = x_h/(1-\sqrt{\Phi})$.

    In \MJ\ the solution is defined between $x = \infty$ (spatial infinity)
    and $x = x_h >1$ (another spatial infinity) and describes a \wh, the
    manifold \MJ\ being geodesically complete. In \ME\ the solution is
    defined in a larger range of $x$, $1 < x < \infty$ where $x=1$ is the
    central singularity. Thus, according to the definition of
    Refs.\,\cite{CC-def}, we have a conformal continuation from \MJ\ to \ME,
    and the transition surface \Str\ in \ME, i.e., the double horizon,
    corresponds to the second spatial infinity in \MJ. It is a novel feature
    since in the previous examples of conformal continuations
    \cite{CC-def,CC-ex} the regular transition surfaces \Str\ were
    obtained by conformal mappings from different kinds of singularities.

    It should be noted that the second spatial infinity in \MJ, $x = x_h$,
    is generally non-flat but rather possesses a solid angle excess or
    deficit, analogously to the well-known feature of global monopoles
    \cite{vil}. Indeed, near $x = x_h$, we have
    $A(x) \approx \half A_2 (x-x_h)^2$, where $A_2 = A''(x_h)$ (the prime
    denotes $d/dx)$. The solid angle deficit $\mu$ (or excess if $\mu < 0$)
    in \MJ\ at $x=x_h$ is determined from the relation
\beq                                                         \label{mu}
     1 - \mu = -\tg^{xx} {\tilde r}{}'{}^2\Big|_{x\to x_h}
	     = \half A_2 (x_h^2 -1),
\eeq
    where ${\tilde r} = \sqrt{-\tg_{\theta\theta}}$ is the radius of a
    coordinate sphere $t=\const,\ x=\const$ in \MJ. Numerical estimates
    for the same parameter values as in (\ref{dub}) give
\beq                                                        \label{solid}
     \begin{array} {lll}
    m = 1,   \ &  A_2 \approx 3.344,    \ & 1-\mu \approx 1.25898, \\[3pt]
    m = 3,   \ &  A_2 \approx 0.234791, \ & 1-\mu \approx 1.02538, \\[3pt]
    m = 10,  \ &  A_2 \approx 0.020105, \ & 1-\mu \approx 1.00246, \\[3pt]
    m = 40,  \ &  A_2 \approx 0.001250, \ & 1-\mu \approx 0.99987.
     \end{array}
\eeq
    (Note that, at large $m$, the solution approaches the \RN\ one in which
    a double horizon occurs at $x_h = m = q$ and a formal substitution to
    (\ref{mu}) leads to $1-\mu = (m^2-1)/m^2 < 1$.) Since, by (\ref{solid}),
    $\mu > 0$ at smaller $m$ and $\mu < 0$ at larger $m$, we conclude by
    continuity that there is (at least one) set of parameters $(m,q)$ for
    which $\mu = 0$ and the \wh\ is twice \asflat: both $m$ and $q$ for it
    must be slightly smaller than 40.

    This confirms the inference of Ref.\,\cite{BrSt07} that the existence of
    such twice \asflat\ \whs\ requires severe fine tuning. Moreover, as
    follows from (\ref{f-wh}) and (\ref{om-wh}), at $x = x_h$ we have in
    \MJ\ an infinite effective gravitational constant $G_{\rm eff}$ since
    (see, e.g., \cite{GPRS06})
\beq
       G_{\rm eff} = \frac{G}{\Phi}\frac{\omega + 2}{\omega + 3/2}
			\sim \Phi^{-2} \quad {\rm as} \quad x\to x_h,
\eeq
    where $G$ is the initial gravitational constant. Thus such \whs, even if
    they exist, cannot connect different parts of our Universe but can only
    be bridges to other universes (if any) with very unusual physics.

\Acknow
{The work was supported by RFBR grant 05-02-17478. KB acknowledges partial
support from FAPESP (Bra\-zil) and thanks for hospitality the colleagues
from the University of S\~ao Paulo, where part of the work was done.
}

\end{document}